\documentclass[prl,amsmath,amssymb,twocolumn]{revtex4}

\usepackage{graphicx}
\usepackage{subfigure}
\usepackage{adjustbox}
\usepackage{bm}
\usepackage{color}
\usepackage{braket}
\usepackage{standalone}
\usepackage{multirow}
\usepackage{tikz}
\usepackage{mathrsfs}
\usepackage{amsmath}
\usepackage{dsfont}
\usepackage[colorlinks,bookmarks=true,citecolor=blue,linkcolor=red,urlcolor=blue]{hyperref}

\newcommand{\bea}{\begin{eqnarray}}
\newcommand{\eea}{\end{eqnarray}}

\begin{document}
\title{Preparing Atomic Topological Quantum Matter by Adiabatic Nonunitary Dynamics}

\author{S. Barbarino$^{1}$, J. Yu$^{2,3}$, P. Zoller$^{2,3}$, J. C. Budich$^{1}$ }

\affiliation{$^1$Institute of Theoretical Physics, Technische Universit\"{a}t Dresden, 01062 Dresden, Germany}
\affiliation{$^2$ Center for Quantum Physics, Faculty of Mathematics, Computer Science and Physics, University of Innsbruck, Innsbruck A-6020, Austria}
\affiliation{$^3$ Institute for Quantum Optics and Quantum Information, Austrian Academy of Sciences, Innsbruck A-6020, Austria}

\begin{abstract}
Motivated by the outstanding challenge of realizing low-temperature states of  quantum matter in synthetic materials, we propose and study an experimentally feasible protocol for preparing topological states such as Chern insulators. By definition, such (non-symmetry protected) topological phases cannot be attained without going through a phase transition in a closed system, largely preventing their preparation in coherent dynamics. To overcome this fundamental caveat, we propose to couple the target system to a conjugate system, so as to prepare a symmetry protected topological phase in an {\emph{extended system}} by intermittently breaking the protecting symmetry. Finally, the decoupled conjugate system is discarded, thus {\emph{projecting}} onto the desired topological state in the target system. By construction, this protocol may be immediately generalized to the class of invertible topological phases, characterized by the existence of an inverse topological order. We illustrate our findings with microscopic simulations on an experimentally realistic Chern insulator model of ultracold fermionic atoms in a {\color{black}{driven spin-dependent}} hexagonal optical lattice. 
\end{abstract}

\maketitle
Recent experiments have reported remarkable progress in realizing synthetic quantum matter with ultracold atoms~\cite{bloch,Dalibard,Bakr,Lin,Weitenberg,Taie2012,pagano2014,StruckPRL,endres,islam,Mazurenko,song}. This includes the engineering of complex Hamiltonians for Chern Insulators ({\color{black}{CIs}}
) in optical lattices~\cite{Aidelsburger,Miyake,atala,Jotzu,Aidelsburger_bis,mancini,stuhl,rem,Wu2016,goldman,cooper}, and the observation of associated non-equilibrium phenomena in quantum dynamics~\cite{wang,flaschner,Tarnowski,Langen,Jinlong,Yi,sun,YI_bis}. Yet, the preparation of low temperature states required for observing the most fascinating phenomena in topological quantum matter has remained a key obstacle. A common approach for gapped quantum phases is adiabatic state preparation, where a low entropy initial state is adiabatically transformed in unitary dynamics to a satisfactory approximation of the desired ground state~\cite{keesling}. However, unitarily evolving a topologically trivial initial state towards a topological phase implies crossing a topological quantum phase transition (TQPT), rendering this approach impractical~\cite{chen,caio,dalessio,privitera,Ulcakar,toniolo}. Below, we address this problem by proposing and studying a robust and widely applicable protocol for preparing paradigmatic topological phases such as CI states~\cite{Haldane,schnyderreview}, as realized with fermionic atoms in a 2D optical lattice. The novel element allowing us to avoid  a TQPT is to complement the target system (S) with a conjugate (duplicate) system (S*). While the joint system adiabatically undergoes unitary evolution, the topological state in S is prepared in a nonunitary fashion, {\color{black}{by}} eventually  removing S* in a dissipative step. 

\begin{figure}[htp!]
	\begin{center}
  	\includegraphics[width=\columnwidth]{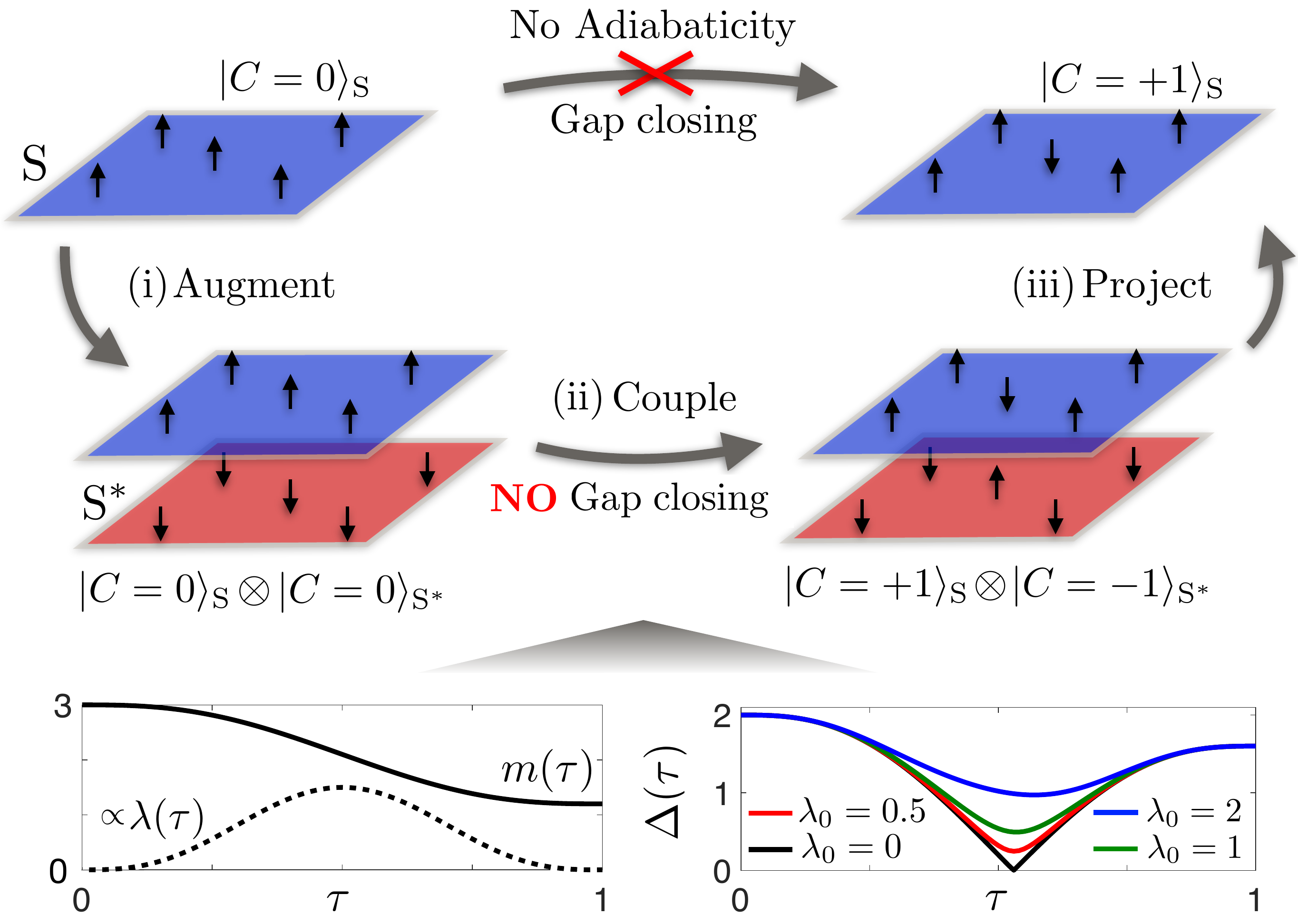}
	\end{center}
	\caption{Protocol for preparing topological states such as Chern insulators (CI) from a trivial initial state in three steps (i)-(iii) (see text). Vertical arrows illustrate the distinction between CI and trivial states. As direct adiabatic preparation of a CI in the target system (S) is impossible, we intermittently couple S to a conjugate  system (S*) so as to keep the total state topologically trivial at all times and avoid a topological quantum phase transition. Finally, S* is projected out to obtain the desired topological phase in S. Insets: data for the model in Eq. (\ref{eqn:dDiracCI}).  
	The S-S* coupling   $\lambda(\tau)=\lambda_0 \tau(1-\tau) \sin^2 \pi \tau$ and the mass term {\color{black}{$m(\tau)= m_0+(m_f-m_0)(10\tau^3-15\tau^4+6\tau^5)$.}}
	 As long as $\lambda_0$ is  finite, the gap $\Delta(\tau)$ stays finite.  
	}
	\label{fig1}
\end{figure} 

Our protocol for preparing a CI state {\color{black}{with Chern number $C$}} in S from the trivial state $\lvert C =0\rangle_\text{S}$ proceeds in three steps (i)-(iii), see Fig.~\ref{fig1}. In step (i), 
S is augmented by S* to 
{\color{black}{form a product state}} $\lvert C=0 \rangle_{\rm S}\otimes\lvert  C=0\rangle_\text{S*}$. This S* is chosen as 
{\color{black}{a conjugated duplicate of S by a symmetry $\mathcal T$, such as time-reversal symmetry (TRS), that reverses the Chern number~\cite{schnyderreview}.}} 
Thus, the composite system has zero Chern number at all times $\tau$ of the  subsequent adiabatic unitary dynamics. The time-dependent 
Hamiltonian of the composite system in reciprocal space at lattice momentum $\boldsymbol k$ is of the form
\begin{align}
h(\boldsymbol k,\tau)=\begin{pmatrix}
h_{\rm S}(\boldsymbol k,\tau)&\Lambda(\tau)\\
\Lambda^\dag(\tau)&h_{\rm S^*}(\boldsymbol k,\tau)
\end{pmatrix},
\label{eqn:totalToyHam}
\end{align}
where lattice translation invariance is assumed and $h_{\rm S}(\boldsymbol k,\tau)$ ($h_{\rm S^*}(\boldsymbol k,\tau)$) is the Hamiltonian of S (S*), and $\Lambda(\tau)$ denotes a time-dependent local coupling between S and S*. In a second step (ii), the combined system {\color{black}{evolves}} to a symmetry protected topological state $\lvert C=+1 \rangle_{\rm S}\otimes\lvert  C=-1\rangle_\text{S*}$ with opposite non-zero Chern {\color{black}{numbers}} 
 in S and S* by tuning a generic parameter $m(\tau)$ in $h_{\rm S}(\boldsymbol k,\tau)$ and $h_{\rm S^*}(\boldsymbol k,\tau)$, respectively. Notably, when intermittently switching on the symmetry breaking coupling  $\Lambda(\tau)$, this can be achieved adiabatically even in the thermodynamic limit, i.e. by maintaining a finite  energy gap $\Delta(\tau)${\color{black}{,}} thus avoiding critical slowdown due to a TQPT. Towards the end of the parameter ramp, S and S* are decoupled (disentangled) again, e.g. by adiabatically switching off  $\Lambda(\tau)$. Then, in a final dissipative step (iii), the {\color{black}{composite}} system is projected onto the target state $\lvert C=+1 \rangle_\text{S}$ by discarding the decoupled S* without affecting the reduced state of S. At a conceptual level, our projective preparation scheme relies on augmenting the desired target state to a topologically trivial total state in an enlarged Hilbert space. Thus it is directly applicable to the class of invertible topological phases~\cite{wen}, i.e. topologically ordered phases for which an inverse topological order in a system of comparable complexity exists, rendering the combined system topologically trivial in the absence of symmetries.
{\color{black}{Invertible topological phases are also of importance for the topological proximity effect~\cite{Hsieh,Zheng,ChengR}, where a topological phase is induced in an initially trivial system by coupling it to a non-trivial state. There, however, the composite system typically becomes topological, in contrast to our present scenario.}}

{\emph{{\color{black}{Square lattice}}  
Chern insulator model.---}}For simplicity, we first explicate the proposed protocol with the help of 
{\color{black}{a minimal CI model defined}}
on a 2D square lattice with unit lattice constant, before applying it to an experimentally feasible model of ultracold atoms \cite{hauke,Sengstock2011,Tarnowski}. In reciprocal space, the model Hamiltonian for S reads 
\begin{align}
h_{\rm S}(\boldsymbol k,\tau)=\mathbf d_{\rm S}(\boldsymbol k,\tau) \cdot \boldsymbol\sigma,
\label{eqn:dDiracCI}
\end{align}
where $\mathbf d_{\rm S}(\boldsymbol k,\tau) = \left(\sin(k_x),\,\sin(k_y),\,m(\tau)-\sum_{i}\cos(k_i)\right)$, $i=x,y$, the mass parameter $m(\tau)$ is ramped with $\tau=t/T\in [0,1]$ with the ramp time $T$, and $\boldsymbol\sigma$ is the vector of 
Pauli matrices. {\color{black}{At half-filling, S has Chern number zero for $\lvert m\rvert >2$, and is in a Chern insulator phase with Chern number $\text{sign}(m)$ for $0<\lvert m\rvert <2$}}. The quantum phase transitions, preventing us from adiabatically preparing a CI phase in S just by changing $m(\tau)$, occur at the critical values $m_c=0,\pm 2$. For S*, we simply choose a TRS conjugated copy of S, i.e. $h_{\rm S^*}(\boldsymbol k,\tau)=h^*_{\rm S}(-\boldsymbol k,\tau)$ in Eq.~(\ref{eqn:totalToyHam}). The coupling between S and S* is assumed as $\Lambda(\tau) = \lambda(\tau) \sigma_x${\color{black}{,}} with a time-dependent coupling strength $\lambda(\tau)$. Using the celebrated SO($5$) Clifford-algebra of $4\times4$ Dirac matrices (see Supplementary Material), it is easy to see that the spectrum of the extended Hamiltonian (\ref{eqn:totalToyHam}) is given by 
\begin{align}
E_\pm(\boldsymbol k,\tau) = \pm \sqrt{\left|\boldsymbol d_{\rm S}( \boldsymbol k,\tau )\right|^2+\lambda^2(\tau)}.
\label{eqn:totalspec}
\end{align}
As a consequence, the gap $\Delta(\tau) = \text{min}_{\boldsymbol k} \,2\lvert E_\pm(\boldsymbol k,\tau)\rvert$ at half-filling of the total system never closes as a function of $\tau$ as long as $\lambda(\tau)$ is finite while $m(\tau)$ passes a critical value $m_c$. {\color{black}{In Fig.~\ref{fig1}, we illustrate the basic steps of this protocol for concrete ramp functions $m(\tau),\,\lambda(\tau)$, chosen such that S and S*,
initially decoupled with zero Chern number, evolve adiabatically with respect to $\Delta(\tau)$ and, at $\tau=1$, reside in a nontrivial CI state with Chern numbers $+1$ and $-1$ respectively, i.e.
$
\lvert C=0 \rangle_{\rm S}\otimes\lvert  C=0\rangle_\text{S*} \xrightarrow{\textrm{(ii)}} \lvert C=+1 \rangle_{\rm S}\otimes\lvert  C=-1\rangle_\text{S*}.
$}}
Finally, the conjugate system S* is discarded, and further experimental manipulation and analysis of the CI state in S may commence.


\begin{figure}
\begin{center}
  \includegraphics[width=\columnwidth]{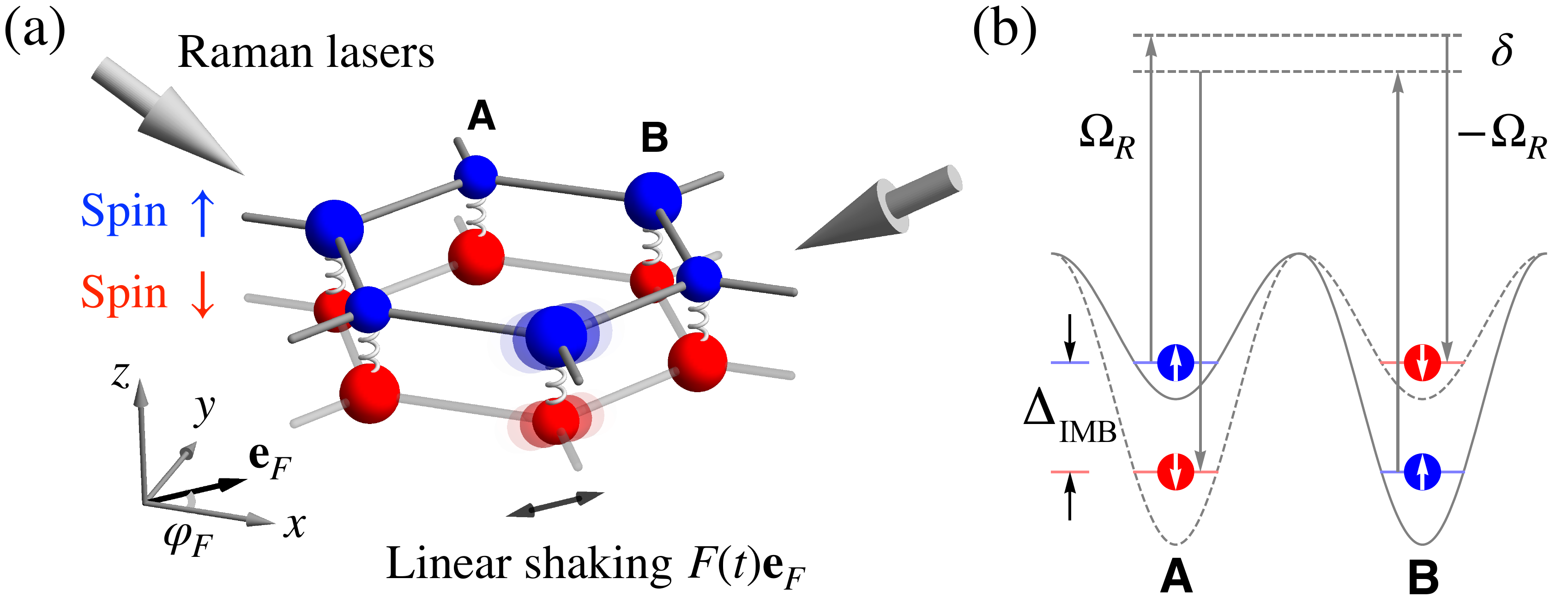}
\end{center}
\caption{(a) A linearly-shaking spin-dependent hexagonal lattice is subject to Raman lasers. The lattice is shaken along the $\mathbf{e}_F$ direction, where $\mathbf{e}_F = \cos \varphi_F \mathbf{e}_x + \sin \varphi_F \mathbf{e}_y$ is a unit vector on the $x$-$y$ plane.  The spin $\uparrow$ and $\downarrow$ sectors---representing the system S and the conjugate system $\rm S^*$, respectively---lay on the same plane, but are artificially lifted  for better visualization. Raman lasers are applied to activate local couplings (as indicated by the springs) between spin $\uparrow$ and $\downarrow$ atoms on the same lattice sites. (b) The spin-dependent optical potentials for spin $\uparrow$ and $\downarrow$ atoms within a unit cell of {\bf A} and {\bf B} sublattice sites. The solid (dashed) curve is the potential for spin $\uparrow$ ($\downarrow$) atoms. Raman lasers with a two-photon detuning $\delta$ flip spins locally. Note that the Raman coupling amplitude ($\pm\Omega_R$) is designed to take a sublattice-dependent sign factor. 
}
\label{Fig:Fig2}
\end{figure}

\begin{figure}[htp!]
  	\includegraphics[width=\columnwidth]{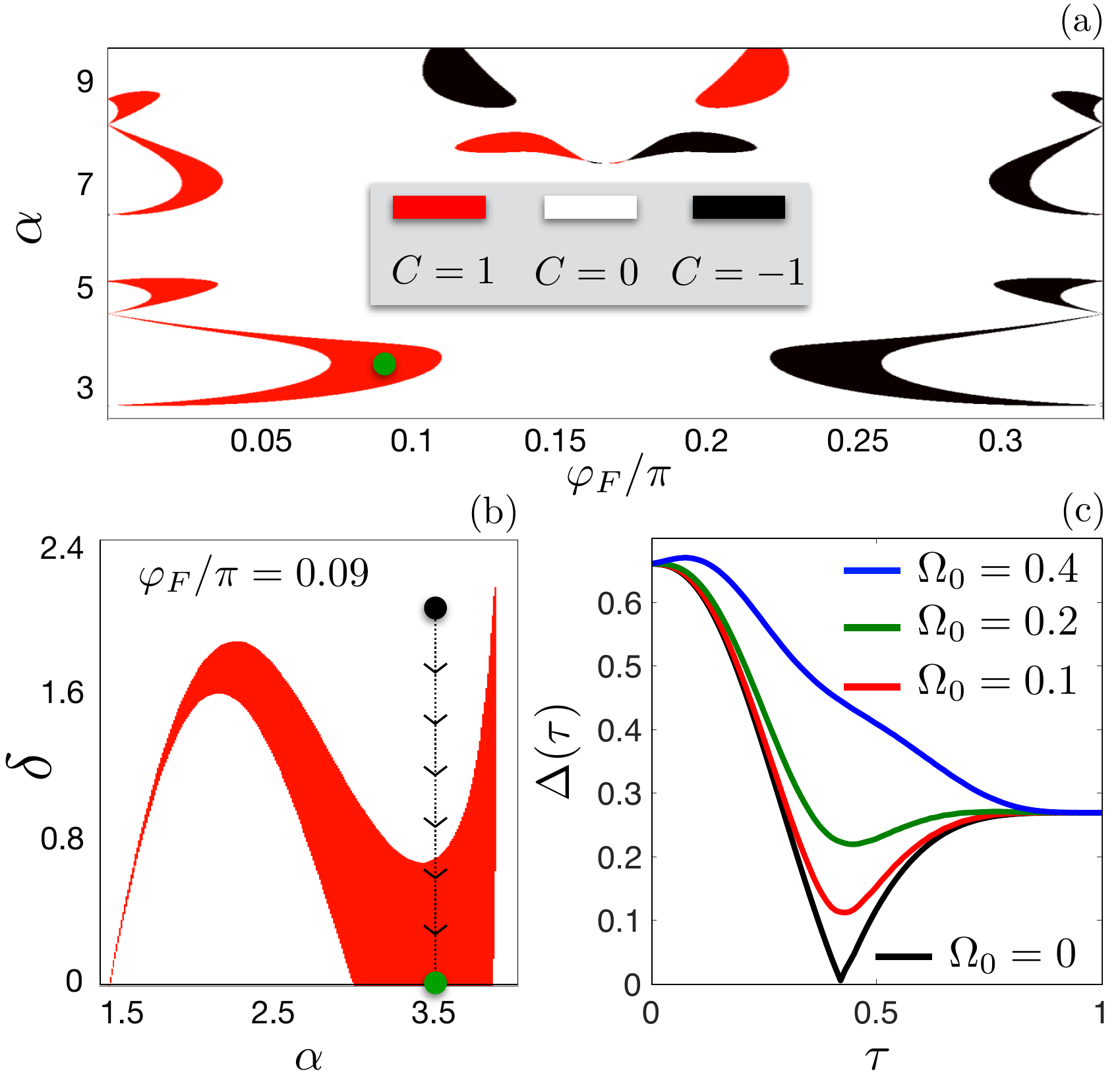}
	\caption{(a) The phase diagram of the system S as a function of $\alpha$ and $\varphi_F$ with $\delta=0$. The conjugate system S* has the opposite Chern number. 
	(b) The phase diagram as a function of $\alpha$ and $\delta$ for $\varphi_F=0.09\pi$.   The (black) green dot indicates the initial (final) state of the ramping protocol. 
	(c) The band gap $\Delta(\tau)$ for different values of the Raman coupling. 
	The ramping protocol for the detuning term is $\delta(\tau)=\delta_i+(\delta_f-\delta_i)f(\tau)$ with  $f(\tau)=15 \tau^7-70\tau^6+126\tau^5-105 \tau^4+35\tau^3$ and  $\tau=t/T$, being $T$ the ramp time.
	The coupling between S and S* is $\Lambda(\tau) =\Omega_R(\tau) \sigma_z$ with $\Omega_R(\tau)=\Omega_0 (1-\tau) \sin \pi \tau$. Here: $t_{\rm NNN}=1$ and $t_{\rm NN}=5$. 
	}
	\label{FIG3}
\end{figure}

{\emph{Implementation with ultracold atoms.---}}We now show how our proposed protocol can be naturally implemented in ultracold atomic gases by combining existing experimental techniques~\cite{hauke,Sengstock2011,Tarnowski}. We consider a spin-dependent hexagonal optical lattice that is linearly-shaken with a characteristic shaking frequency $\omega_D$, and subjected to a set of Raman lasers with a Raman frequency $\omega_R$ [see Fig.~\ref{Fig:Fig2}(a)], where we choose $\omega_D = \omega_R$ for simplicity. In reciprocal space, in the absence of the Raman coupling,  the system is governed  by a time-independent effective Hamiltonian $H_t(\boldsymbol{k}) + H_\text{IMB}(\boldsymbol{k})$. Here,  $H_t(\boldsymbol{k})$ describes the hopping of atoms as \cite{hauke,SSMM}
\begin{equation} \label{Eq:H_hopping}
\begin{aligned}
H_t(\boldsymbol{k}) =&\sum_{s=\uparrow,\downarrow} [g_{\alpha, \varphi_F}(\boldsymbol{k}) a^\dagger_{\boldsymbol{k}s} b_{\boldsymbol{k}s} + \mathrm{h.c.}] \, + \\
&+g'_{\alpha, \varphi_F}(\boldsymbol{k}) (a^\dagger_{\boldsymbol{k}\uparrow} a_{\boldsymbol{k}\uparrow} + b^\dagger_{\boldsymbol{k}\downarrow} b_{\boldsymbol{k}\downarrow}),
\end{aligned}
\end{equation}
where $a_{\boldsymbol{k}s}$ ($b_{\boldsymbol{k}s}$) is the annihilation operator for a fermion with spin $s$ in the sublattice {\bf A} ({\bf B}). The structure function  $g_{\alpha, \varphi_F}(\boldsymbol{k})$ ($g'_{\alpha, \varphi_F}(\boldsymbol{k})$)  is proportional to the bare nearest-neighbor $t_{\rm NN}$  (next-nearest-neighbor $t_{\rm NNN}$) hopping coefficients and depends~\cite{SSMM} on the shaking strength $\alpha$ {\color{black}{proportional to the amplitude of the linear shaking}}  as well as on the shaking direction $\varphi_F$, see Fig.~\ref{Fig:Fig2}(a).
The second term $H_{\rm IMB}(\boldsymbol{k})$ describes an effective spin-dependent sub-lattice imbalance~\cite{SSMM}
\begin{equation}
H_{\rm IMB}(\boldsymbol{k})= \frac{\delta}{2}  ( {a_{ \boldsymbol{k}\uparrow }^\dag {a_{\boldsymbol{k} \uparrow }} - a_{\boldsymbol{k} \downarrow }^\dag {a_{\boldsymbol{k} \downarrow }}} -{b_{\boldsymbol{k} \uparrow }^\dag {b_{\boldsymbol{k} \uparrow }} +b_{\boldsymbol{k} \downarrow }^\dag {b_{\boldsymbol{k}\downarrow }}}) \,,
\label{det}
\end{equation}  
where $\delta = \Delta_\text{IMB} - \omega_D$ is the detuning between the sublattice imbalance $\Delta_\text{IMB}$ and the shaking frequency $\omega_D$~\cite{omegaD}.
We observe that the Hamiltonian $H_t(\boldsymbol{k})+H_{\rm IMB}(\boldsymbol{k})$ can be recast in the form of Eq.~\eqref{eqn:totalToyHam}, so far without 
the coupling $\Lambda$ between S and S*, where S and S* are identified with the spin $\uparrow$ and $\downarrow$ sectors, respectively. 
Introducing the vector $\mathbf d(\boldsymbol k)$ as
\begin{subequations}
\begin{align}
&{d_x}({\boldsymbol{k}}) = \operatorname{Re} \left[ {g_{\alpha, \varphi_F}({\boldsymbol{k}})} \right], \\
&{d_y}({\boldsymbol{k}}) =  - \operatorname{Im} \left[ {g_{\alpha, \varphi_F}({\boldsymbol{k}})} \right], \\
&{d_z}({\boldsymbol{k}}) =  [\delta + g'_{\alpha, \varphi_F}({\boldsymbol{k}})] /2 \, ,
\end{align}
\end{subequations}
the Hamiltonian  ${h_{\rm S}}(\boldsymbol{k})$ describing S can be written \textcolor{black}{as}  ${h_{\rm S}}(\boldsymbol{k})=({d_x},{d_y},{d_z}) \cdot \bm{\sigma}$,
while ${h_{\rm S^*}}(\boldsymbol{k}) = ({d_x},{d_y}, - {d_z}) \cdot \bm{\sigma}$ characterizes the S*. This close relation between $h_{\rm S}(\boldsymbol{k}) $ and $h_{\rm S^*}(\boldsymbol{k}) $, playing a similar role as the standard TRS in the aforementioned minimal model, is crucial for our protocol: It manifestly renders the Chern numbers of S and S* opposite, such that the combined system always remains topologically trivial~\cite{SSMM}. For the case of  $\delta = 0$, the phase diagram of S is shown in Fig.~\ref{FIG3}(a) as a function of $\alpha$ and $\varphi_F$, from which topological CI regimes characterized by a nonzero Chern number can be identified. 
When considering finite detuning, $\delta(\tau)$  can be used to tune the topology of S between different phases [see Fig.~\ref{FIG3}(b)], thus playing a similar role to the mass parameter $m(\tau)$ in the aforementioned toy model (\ref{eqn:dDiracCI}).
We note that, the above Hamiltonian $H_t(\boldsymbol{k})+H_{\rm IMB}(\boldsymbol{k})$ can be directly realized in the experimental setup of Ref.~\cite{Tarnowski}, by putting a second spin species onto the lattice (as in Ref.~\cite{Sengstock2011}) and replacing the circular shaking with a linear one. 

When the Raman lasers are turned on, the S-S* coupling $\Lambda(\tau)$ as in Eq.~\eqref{eqn:totalToyHam} is activated.
If we arrange the Raman lasers such that the phase factors for sublattices {\bf A} and {\bf B} take an opposite sign,
within the rotating-wave approximation, the inter-spin Raman coupling is described by the time-independent Hamiltonian~\cite{SSMM}
\begin{equation}
H_R(\boldsymbol{k}) ={\Omega _R} {\left( {a_{\boldsymbol{k}\uparrow }^\dag {a_{\boldsymbol{k} \downarrow }}-b_{\boldsymbol{k} \uparrow }^\dag {b_{\boldsymbol{k} \downarrow }}+ {\text{h.c.}}} \right)},
\end{equation}
which engineers the coupling term  $\Lambda = {\Omega _R}{\sigma_z}$ in Eq.~\eqref{eqn:totalToyHam}. 

With all the ingredients in place, our projective preparation protocol can be summarized as follows. 
A topologically non-trivial target state of S, e.g., the green dot in Fig.~\ref{FIG3}(b) in the Chern number $+1$ phase, can be adiabatically reached within a sufficiently large (but system-size independent) ramp time $T$ by reducing $\delta(\tau)$ from an initial value $\delta(\tau=0)=\delta_i$ to the desired final value $\delta(\tau=1)=\delta_f$ at fixed $\alpha$ and $\varphi_F$ [as illustrated in Fig.~\ref{FIG3}(b)], while the inter-spin (S-S*) coupling $\Lambda(\tau) =\Omega_R(\tau) \sigma_z$ is switched on (off) at the beginning (end) of the protocol. During this parameter ramp, the spectrum is again of the form (\ref{eqn:totalspec}) with $\lambda(\tau)=\Omega_R(\tau)$~\cite{SSMM}. Thus, analogous to the minimal model illustrated in Fig.~\ref{fig1}, the local Raman coupling $\Omega_R(\tau)$ is indeed sufficient to maintain a finite gap $\Delta(\tau)$ at all times [see Fig.~\ref{FIG3}(c)]. As a final step of the protocol, we project the adiabatically decoupled total system onto the target system S by selectively removing S* using, e.g., a magnetic field gradient~\cite{Jotzu}.

{\emph{Microscopic simulation of parameter quench.---}}
In order to quantitatively assess the practical applicability {\color{black}{of}} our protocol to the ultracold atomic gas model derived in the previous paragraph,  
we now explicitly simulate the quench-dynamics of the system from a trivial initial state towards the topological target state. Concretely, we assume $\alpha=3.5$, $\varphi_F=0.09\pi$ {\color{black}{such that, at the end of the protocol, the system S has Chern number $+1$, see also Fig.~\ref{FIG3}(a)}}. The initial detuning $\delta_i=2$ is progressively reduced to the final value $\delta_f=0$,  while  the S-S* coupling strength is varied according to $\Omega_R(\tau)=\Omega_0 (1-\tau) \sin \pi \tau$ with $\Omega_0=0.4$, such that the minimum gap encountered during the total time evolution is given by the final gap of the model, i.e. $\Delta(T) \approx 0.27$; see Fig.~\ref{FIG3}(c).

\begin{figure}
  	\includegraphics[width=\columnwidth]{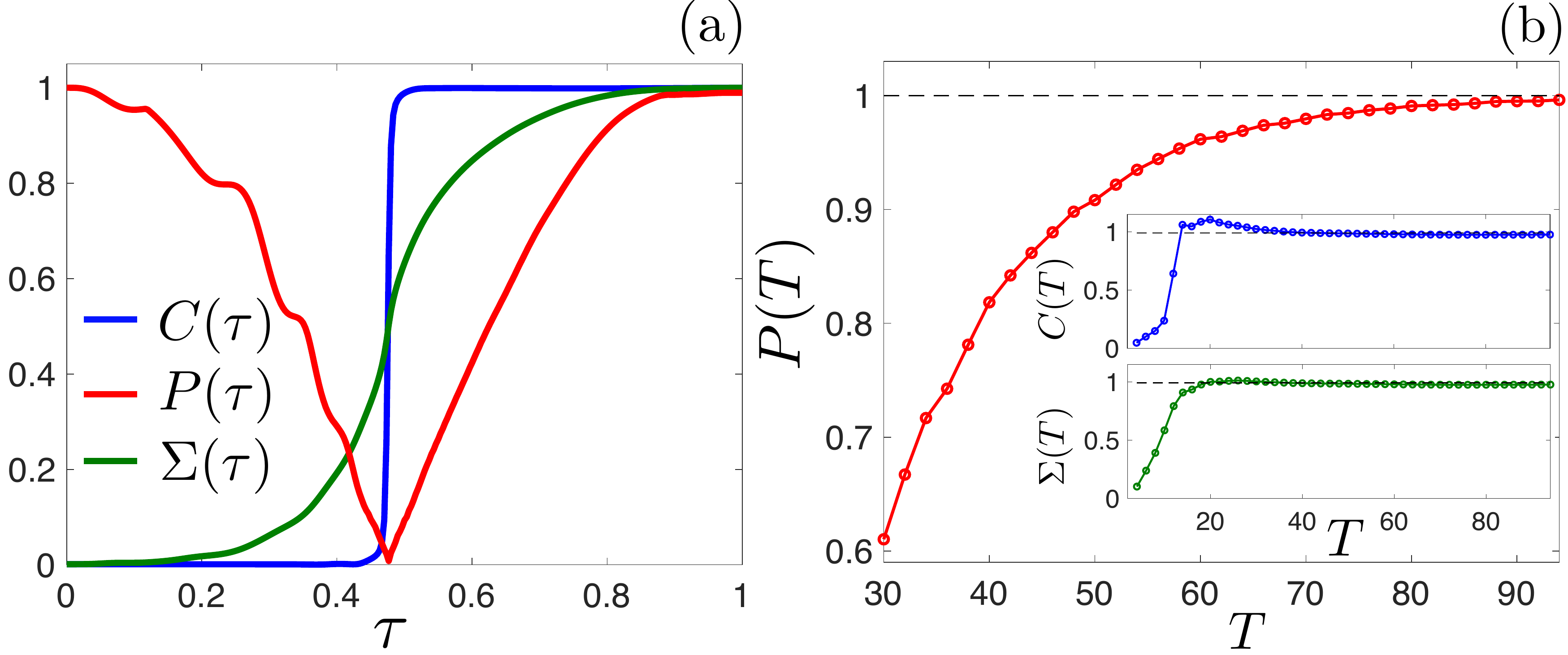}
	\caption{(a) Time evolution of the Chern number, the Hall conductivity and the minimum of the purity for $T=80$. 
	(b) The minimum of the purity gap $P(T)$ as a function of the ramping time $T$. Insets: the Chern number $C(T)$ and the Hall conductivity $\Sigma(T)$ as a function  of the ramping time $T$; here $\varphi_F=0.09\pi$, $\alpha=3.5$,
	$\delta_i=2$, $\delta_f=0$ and $\Omega_0=0.4$.  
		}
	\label{FIG4}
\end{figure}

Since we are eventually interested in the properties of the target S only, we consider its reduced density matrix 
$\rho_{{\boldsymbol k}}(\tau)=\frac{1}{2}(1+ \boldsymbol{ \tilde d}_{\boldsymbol k}(\tau) \cdot \boldsymbol{\sigma})$, where the Bloch vector $\boldsymbol {\tilde d}_{\boldsymbol k}(\tau)$ describes the polarization of $\rho_{{\boldsymbol k}}(\tau)$ on the Bloch sphere. 
Its length $p_{\boldsymbol k}(\tau)=|\boldsymbol {\tilde d}_{\boldsymbol k}(\tau)| \leq 1$, known as purity gap~\cite{diehl,hu,budich}, measures the purity of the state associated to $\rho_{\boldsymbol k}(\tau)$. 
In Fig.~\ref{FIG4}(a), we show the minimum of the purity gap, i.e. $P(\tau)=\min_{{\boldsymbol k}}[ p_{\boldsymbol k}(\tau)]$, during the quench-dynamics.  
For the shown data where $T \gg [\Delta(T)]^{-1}$, the purity gap saturates close to one at the end of the ramp confirming that a topologically non-trivial target state of S can be adiabatically reached within a sufficiently large ramp time $T$. 
In order to probe the topological nature of S, we also monitor the time-dependent Chern number~\cite{hu}  
\begin{align}
C(\tau)=\frac{1}{2\pi} \int_{\rm BZ} d {\boldsymbol k} \, \mathcal{F}_{\boldsymbol k}(\tau)
\label{eqn:Chernoft}
\end{align}
and the instantaneous equilibrium Hall conductance 
\begin{align}
\Sigma(\tau)=\frac{1}{2\pi} \int_{\rm BZ} d {\boldsymbol k} \, \mathcal{F}_{\boldsymbol k}(\tau) \, p_{\boldsymbol k}(\tau)
\label{eqn:Hall}
\end{align}
where  $\mathcal{F}_{\boldsymbol k}(\tau)=-\frac 12 \boldsymbol n_{\boldsymbol k}(\tau) \cdot [ \partial_{k_x} \boldsymbol n_{\boldsymbol k}(\tau) \times  \partial_{k_y} \boldsymbol n_k(\tau) ] $ is the Berry curvature, $\boldsymbol n_{\boldsymbol k}(\tau) \equiv \boldsymbol {\tilde d}_{\boldsymbol k}(\tau) /p_{\boldsymbol k}(\tau)$ and BZ is the Brillouin zone.  In Fig.~\ref{FIG4}(a), we  display $C(\tau)$ and $\Sigma(\tau)$, noting that the Chern number exhibits a sudden jump when the minimum of the purity goes to zero, while $\Sigma(\tau)$ saturates smoothly. Finally, to quantitatively assess deviations from adiabaticity due to finite $T$, in Fig.~\ref{FIG4}(b), we study the minimum of the purity gap at the end of the protocol, i.e. $P(T)$, for different values of the ramp time $T$, where perfect adiabaticity reflected in  $P(T) \sim 1$ is found when $T \gg [\Delta(T)]^{-1}$.
From Eq.~(\ref{eqn:Hall}), it is clear that $P(T)$ directly bounds deviations from the quantized Hall conductance in the post-quench steady state.
In the two insets of Fig.~\ref{FIG4}(b), we show the Chern number and the Hall conductance as a function of $T$ and verified that they both converge to the quantized value of the target Chern state upon increasing $T$. 

In our simulations, we have taken $t_{\rm NNN} = 1$ as energy unit, and set $t_{\rm NN} = 5$ throughout. For a typical experimental system, e.g., as in Ref.~\cite{Tarnowski}, these parameters are given as $t_{\rm NNN} \sim (2\pi\times) 100\, {\rm Hz}$ and  $t_{\rm NN} \sim (2\pi\times) 500\, {\rm Hz}$. Thus, the time unit of Fig.~\ref{FIG4}(b) corresponds to $\sim 10 \,{\rm ms}$. We see that, at the end of our protocol, the Chern number and the Hall conductance are well quantized to one for a preparation time $T \sim 30$, or $300 \,{\rm ms}$, which is close to the typical state-preparation time in experiments (see, e.g., Ref~\cite{Jotzu}). To reach the ideal value $P \sim 1$ for the purity gap, it takes more time ($T\sim80$, or $800 \,{\rm ms}$), which should also be well within reach for state-of-the-art experimental systems.

{\emph{Concluding discussion.---}}Above, we illustrated our preparation protocol for topological states with an experimentally feasible two-banded CI model of ultracold atoms in a hexagonal optical. It is natural to ask as to what extent this concrete scheme may be generalized to other microscopic models or even different topological phases. Regarding alternative CI models such as the experimentally studied Hofstadter model~\cite{Aidelsburger,Miyake,Aidelsburger_bis}, we note that a {\emph{local one-body}} coupling $\Lambda(\tau)$ between S and S* can always be found so as to maintain a finite gap during the entire preparation, and S* may always be chosen to duplicate the degrees of freedom of S. However, as a practical complication, for models with a larger (magnetic) unit cell, a spatial dependence of the coupling $\Lambda(\tau)$ within the unit-cell may become necessary. Turning to different topological phases, the class of {\emph{invertible}} topological phases~\cite{wen}, e.g. encompassing also topological superconductors~\cite{qizhang}, basically by definition meets the relevant criteria for the protocol proposed in this work: Invertible topological phases are defined by the existence of an inverse topological order which complements a given topological order to a trivial total state. As for the particularly intriguing class of fractional quantum Hall (FQH) states, augmenting a target S by an S* consisting of a time-reversal conjugated copy of S always leads to zero Hall conductance. However, depending on the specific FQH phase under consideration, the combined system thus defined may still be either topologically ordered, or connected to a trivial state~\cite{stern}. Thus, a direct generalization of our protocol may only be within reach for certain FQH states. The general question as to what minimal complexity of the S* is required to augment any topological state in S to a trivial total state is an interesting subject of future work.

In our calculations, we assumed periodic boundary conditions to simulate the bulk of large translation invariant systems towards the thermodynamic limit. In experiments with open boundaries, the target CI state features gapless chiral edge states. To reduce the occurrence of edge excitations during the state preparation, it is thus desirable that the coupling $\Lambda(\tau)$ also creates a finite gap on the edge of the system. For both examples studied in our present work, we carefully verified that an edge gap proportional to $\Lambda(\tau)$ is indeed present at all times during the preparation process, leading to the formation of gapless edge states only in the final target state on switching off $\Lambda(\tau)$~\cite{SSMM}.

\acknowledgments
{\it Acknowledgments.---}J.C.B. acknowledges financial support from the German Research Foundation (DFG) through the Collaborative Research Centre SFB 1143 (project-id 247310070) and the W\"urzburg-Dresden Cluster of Excellence on Complexity and Topology in Quantum Matter – ct.qmat (EXC 2147, project-id 39085490).
S.B. acknowledges the Hallwachs-R\"ontgen Postdoc Program of ct.qmat for financial support. 
Work at Innsbruck was supported by PASQuanS, EU Quantum Flagship, QTFLAG – QuantERA, and the Simons foundation via the Simons collaboration UQM.
We thank M. Dalmonte, L. Pastori, and C. Repellin for fruitful discussions.

\clearpage
\title{Supplemental Material for: \\
Preparing Atomic Topological Quantum Matter by Adiabatic Nonunitary Dynamics
}

\onecolumngrid
\setcounter{figure}{0}
\setcounter{equation}{0}
\renewcommand\thefigure{S\arabic{figure}}
\renewcommand\theequation{S\arabic{equation}}
\renewcommand{\bibnumfmt}[1]{[S#1]}
\renewcommand{\citenumfont}[1]{S#1}

\subsection{Details for the square lattice
Chern insulator model}
\subsubsection{Energy spectrum }
We consider the Hamiltonian of {\color{black}{the minimal CI model defined on a 2D square lattice}} 
 \begin{align}\label{Eq:Hami_Coupled_QAH}
h(\boldsymbol k,\tau)=\begin{pmatrix}
h_{\rm S}(\boldsymbol k,\tau)&\Lambda(\tau)\\
\Lambda^\dag(\tau)&h_{\rm S^*}(\boldsymbol k,\tau)
\end{pmatrix},
\end{align}
where $h_{\rm S}(\boldsymbol k,\tau)=\mathbf d_{\rm S}(\boldsymbol k,\tau) \cdot \boldsymbol\sigma$  with $\boldsymbol\sigma=(\sigma_x, \sigma_y, \sigma_z)$ being the vector of Pauli matrices, and we denote $\mathbf d_{\rm S}(\boldsymbol k,\tau)=(d_x(\boldsymbol k), d_y(\boldsymbol k), d_z(\boldsymbol k,\tau))$ with 
 \begin{subequations}
 \begin{align}
 d_x(\boldsymbol k)&=\sin(k_x) ,\\
 d_y(\boldsymbol k)&=\sin(k_y) ,\\
 d_z(\boldsymbol k,\tau)&= m(\tau)-\cos(k_x)-\cos(k_y) .
 \end{align}
 \end{subequations}
Furthermore, we consider the case that $\rm S^*$ is a time-reversal conjugate of S: $h_{\rm S^*}(\boldsymbol k,\tau)=h^*_{\rm S}(-\boldsymbol k,\tau)$, which corresponds to $\mathbf d_{\rm S^*}(\boldsymbol k,\tau) =\mathbf d_{\rm S}(-\boldsymbol k,\tau)$. The coupling between S and $\rm S^*$  is chosen to take the form $\Lambda(\tau) = \lambda(\tau) \sigma_x$. The function $\lambda(\tau)$ is chosen to satisfy $\lambda(\tau = 0)=\lambda(\tau=1)=0$. 
If we now introduce another set of Pauli matrices $s_x$, $s_y$, $s_z$, and the two-by-two identity matrix $s_0$ to describe the S-$\rm S^*$ degree-of-freedom, we can define the following five anti-commuting 4$\times$4 Dirac matrices $\Gamma_\mu$ ($\mu=1,\dots,5$):
\begin{equation} \label{Eq:Gamma_matrices_QAH}
	 \Gamma_1= s_z\otimes\sigma_x,\; \Gamma_2=s_0\otimes\sigma_y,\; \Gamma_3=s_0\otimes \sigma_z,\;
 \Gamma_4= s_x\otimes\sigma_x,\; \Gamma_5=s_y\otimes\sigma_x,
\end{equation}
and rewrite the Hamiltonian $h(\boldsymbol k,\tau)$ as follows:
 \begin{align} \label{Eq:h_k_tau_gamma-matrices}
h(\boldsymbol k,\tau)= d_x(\boldsymbol k) \Gamma_1 + d_y(\boldsymbol k) \Gamma_2 + d_z(\boldsymbol k,\tau) \Gamma_3 + \lambda(\tau) \Gamma_4 \,.
\end{align}
Taking into account that the matrices $\Gamma_\mu$ satisfy the anti-commuting $SO(5)$ Clifford algebra
\begin{align}
\left\{\Gamma_\mu,\Gamma_\nu\right\}=2\delta_{\mu\nu},
\label{eqn:cliff5}
\end{align}
we can calculate the spectrum
\begin{align}
E_\pm(\boldsymbol k,\tau) = \pm \sqrt{|\boldsymbol d_{\rm S}(\boldsymbol k,\tau)|^2+\lambda^2(\tau) } \, ,
\label{non_int_spectrum}
\end{align}
which corresponds to Eq. (3) in the main text. The spectral gap is then given by $\Delta(\tau)= 2\,\text{min}_{\boldsymbol k} [ E_+(\boldsymbol k,\tau) ]$, which is generally nonzero for all the parameters between $\tau = 0\to1$. 

\subsubsection{Edge modes during state preparation}
\begin{figure}[htp!]
	\begin{center}
  	\includegraphics[width=\columnwidth]{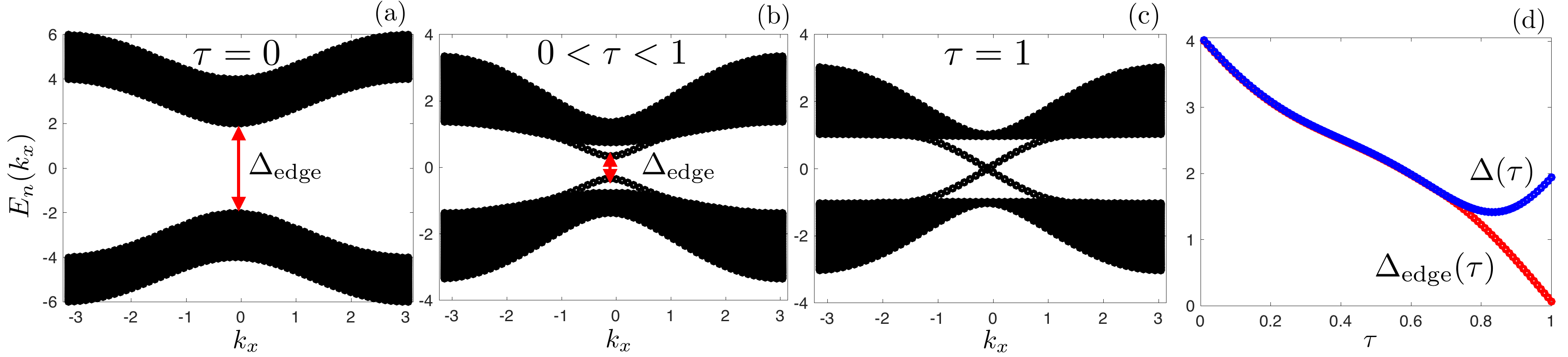}
	\end{center}
	\caption{Energy spectrum properties for the square lattice Chern insulator model with open boundary conditions along the $y$ direction. (a)-(c) The noninteracting spectrum for different values of $\tau$. (d) The gap $\Delta(\tau)$ with periodic boundary conditions and the gap $\Delta_{\rm edge}(\tau)$ with open boundary conditions as a function of the ramp time $\tau$. Simulations are performed with the following parameters: $m(\tau)=m_0+(m_f-m_0)f(\tau)$ with $f(\tau)=10 \tau^3-15\tau^4+6\tau^5$, $m_0=3$, $m_f=1$, and $\lambda(\tau)=\lambda_0 \tau(1-\tau) \sin^2 \pi \tau$ with $\lambda_0=1$. The number of sites along the $y$ direction is $L=60$.
	}
	\label{fig_sm}
\end{figure}

In experiments with open boundaries, the target Chern insulator state features gapless chiral edge states. To reduce the occurrence of edge excitations during the state preparation, it is thus desirable that the coupling $\lambda(\tau)$ also creates a
finite gap on the edge of the system. In the following, we show that an edge gap proportional to $\lambda(\tau)$ is indeed present at all times during the preparation process, leading to the formation of gapless edge states only in the final target state when the coupling $\lambda$ is switched off. To this aim, we consider open boundary conditions along the $y$ direction (and periodic boundary conditions along the $x$ direction). The corresponding Hamiltonian can be obtained by considering $h(\boldsymbol k,\tau)$ and performing a Fourier transform to real space along the $y$ direction only. As expected,  when open boundary conditions are assumed, edge modes appear and the non-interacting spectrum is not given by Eq. \eqref{non_int_spectrum} anymore. In Fig. \ref{fig_sm}(a-c), we show the non-interacting spectrum for different values of $\tau$.  We observe that during the time evolution, i.e., $0<\tau<1$, some intra-gap modes start to appear and they result in the gapless chiral edge states of the target CI state, see Figs. \ref{fig_sm}(b-c). For each momentum $k_x$, we get the eigenvalues $E_n(k_x)$ with $n=1,2, \dots, 2L$, where $L$ the number of sites along the $y$ direction. 
We can then define the edge gap as 
\begin{equation}
\Delta_{\rm edge}(\tau)= \text{min} |E_{L+2}(k_x)-E_{L}(k_x)|, 
\label{edgeee}
\end{equation}
which corresponds to the energy difference as indicated in Fig. \ref{fig_sm}(b).  As shown in  Fig. \ref{fig_sm}(d), the gap $\Delta_{\rm edge}(\tau)$ is finite  at all times during the preparation process and vanishes at the end of the ramp leading to the formation of gapless edge states only in the final target state.

\subsection{Details for the system to be implemented with ultracold atoms}
\subsubsection{Linearly-shaken spin-dependent hexagonal optical lattice}
In the following, we describe the Hamiltonian for the linearly-shaken  spin- ($s=\uparrow, \, \downarrow$) dependent hexagonal (with sublattice indexes $\eta = {\rm \bf A, B}$) optical lattice~\cite{haukee}: $ H_b+H_{\rm IMB}+H_D(t)$, where, as elaborated below, $H_b$ is the hopping term, $H_{\rm IMB}$ is the sublattice imbalance term, and $H_D(t)$ is the linear-shaking term. 
As discussed in the main text, the $\uparrow$ and $\downarrow$ spin sectors play the role of the system S and the conjugate system $\rm S^*$, respectively. 

The bare hopping Hamiltonian ${H}_b$ consists of spin-dependent tunneling terms
between nearest-neighbor (NN) sites and next-nearest-neighbor (NNN) sites and is described by the Hamiltonian
\begin{equation} \label{Eq:H_bare}
\begin{aligned}
{{ H}_b} &= \sum\limits_{ {i,j} } \sum\limits_{s=\uparrow, \, \downarrow }  {{t_{ij, s }}c_{i,s }^\dag {c_{j,s }}} \, ,  \\
\end{aligned}
\end{equation}
where  $c_{i,s}$ is the annihilation operator for a fermion with spin $s$ on site $i$, and $t_{ij,s}$ is the (spin-preserving) hopping amplitude between sites $i$ and $j$ for the spin component $s$. The summation over $i$ and $j$ is restricted to NN and NNN sites. We also consider the presence of a spin-dependent sublattice imbalance
\begin{equation} \label{Eq:H_IMB}
\begin{aligned}
H_{\rm IMB} &= \frac{\Delta_{\rm IMB}}{2}\sum\limits_{j } \left[ {\left( {a_{j \uparrow }^\dag {a_{j \uparrow }} - a_{j \downarrow }^\dag {a_{j \downarrow }}} \right)}  - {\left( {b_{j \uparrow }^\dag {b_{j \uparrow }} - b_{j \downarrow }^\dag {b_{j \downarrow }}} \right)} \right],
\end{aligned}
\end{equation}
where we have denoted ${a_{j,s }} \equiv  {\left( {{c_{j,s }}} \right)_{j \in \rm \bf A}}$ and ${b_{j,s }} \equiv  {\left( {{c_{j,s }}} \right)_{j \in \rm \bf B}}$.
In addition, the lattice is subject to a spin-independent lattice shaking
\begin{equation} \label{Eq:H_D_t}
	{{ H}_D}(t) =  \sum\limits_{i,s } {\left( {{\mathbf{F}}(t) \cdot {{\mathbf{r}}_i}} \right)} \, c_{i,s }^\dag {c_{i,s }},
\end{equation}
generated by a unidirectional driving force ${\mathbf{F}}(t) = F(t){{\boldsymbol{e}}_F}$ with ${{\boldsymbol{e}}_F} = \cos ({\varphi _F}){{\boldsymbol{e}}_x} + \sin ({\varphi _F}){{\boldsymbol{e}}_y}$, which has a paused-sine-wave amplitude~\cite{StruckPRL1} 
\begin{equation}
	F(t) = \left\{ \begin{aligned}
  &{F_0}\sin (\omega_D t),\quad & 0 &\leqslant t\,\bmod \,T_D < {T_1}, \\
  &0,\quad & {T_1}& \leqslant t\,\bmod \,T_D < T_D.
\end{aligned}  \right.  
\end{equation}
Here, $T_D$ is the driving period and $\omega_D = 2\pi/T_1$. 

We now transform the Hamiltonian $ H_b+H_{\rm IMB}+H_D(t)$ into another frame to nullify the driving term $H_D(t)$, i.e., $ H_b+H_{\rm IMB}+H_D(t) \to  H' _b+H'_{\rm IMB}= U_{D}^\dag  HU_D - iU_{D}^\dag {\partial _t}U_D$, with
\begin{equation} \label{Eq:U_D_t}
	{{ U}_D}(t) = \exp \left[ -{i\sum\limits_j {\left( {{\mathbf{A}}(t) \cdot {{\mathbf{r}}_j}} \right)c_{j,s }^\dag {c_{j,s }}} } \right],
\end{equation}
where
\begin{equation}
	{\mathbf{A}}(t) = \int_0^t {{\mathbf{F}}(\tau )d\tau }  - {\left\langle {\int_0^t {{\mathbf{F}}(\tau )d\tau } } \right\rangle _{T_D}}
\end{equation}
is a time-dependent vector field due to the driving and we have denoted ${\left\langle  \cdot  \right\rangle _{T_D}} = \frac{1}{T_D}\int_0^{T_D} { \cdot dt} $. 
As a result of the unitary transformation $U_D$, the fermionic operator ${c_{j,s }}$ gets a time-dependent phase factor
\begin{equation}
	U_D^\dag {c_{j,s }}{U_D} = e^ { - i{\mathbf{A}}(t) \cdot {{\mathbf{r}}_j}} {c_{j,s }} ,
\end{equation}
and the bare Hamiltonian becomes
\begin{equation}
	{{ H}_b^\prime} = \sum\limits_{ {i,j} } \sum\limits_{s=\uparrow, \downarrow} {\left( {{t_{ij,s }}{e^{i{\mathbf{A}}(t) \cdot {{\mathbf{r}}_{ij}}}}} \right)c_{i,s }^\dag {c_{j,s }}} .
\end{equation}
Here, the double-index short-hand $\mathbf{r}_i - \mathbf{r}_j \equiv \mathbf{r}_{ij}$ is adopted. The sublattice imbalance $H_{\rm IMB}$ is unaffected by this unitary transformation, i.e., $H'_{\rm IMB}=U^\dagger_D(t)H_{\rm IMB}U_D(t)=H_{\rm IMB}$. 

\subsubsection{Adding Raman coupling}
We now introduce a Raman coupling term which implements the coupling between the system S and the conjugate system $\rm S^*$. 
In the continuous space, it can be written as
\begin{equation}
	{V_R}({\mathbf{r}},t) = \frac{{V_R^0}}{2}\cos ({\mathbf{q}} \cdot {\mathbf{r}} - {\omega _R}t + \phi_R ){s _x},
\end{equation}
where $V_R^0$ is the amplitude of the coupling, $\mathbf{q}$ and $\omega_R$ are the momentum and frequency difference of the two Raman lasers, respectively, $\phi_R$ is a static phase factor, and $s_x$ is the first Pauli matrix in the spin space which flips the spin from up to down and vice-versa. The fermionic field operator can be expanded  in the real space as 
\begin{equation}
	 \psi ({\mathbf{r}},t) = \sum\limits_{j,s } {\left[ {{w_{{\rm \bf A},j,s }}({\mathbf{r}},t){a_{j,s }} + {w_{{\rm \bf B},j,s }}({\mathbf{r}},t){b_{j,s }}} \right]} ,
\end{equation}
where ${{w_{{\rm \bf A},j,s }}({\mathbf{r}},t)}$ and ${{w_{{\rm \bf B},j,s }}({\mathbf{r}},t)}$ are the Wannier wavefunctions for {\bf A} and {\bf B} sublattices at site $j$, respectively. The Raman coupling term in the second-quantized form is then given by 
\begin{equation}
	\begin{aligned}
  {{ H}_R}(t) =& \int {d{\mathbf{r}}} {{ \psi }^\dag }({\mathbf{r}},t){V_R}({\mathbf{r}},t) \psi ({\mathbf{r}},t) \hfill \\
   =& \frac{{V_R^0}}{2}\sum_j\left[ {\left( {\int {d{\mathbf{r}}} w_{{\rm \bf A},j \uparrow }^*({\mathbf{r}},t){w_{{\rm \bf A},j \downarrow }}({\mathbf{r}},t)\cos ({\mathbf{q}} \cdot {\mathbf{r}} - {\omega _R}t + \phi_R )a_{j \uparrow }^\dag {a_{j \downarrow }} + {\text{h.c.}}} \right) + (a \to b)} \right]. \hfill \\
\end{aligned}
\end{equation}
We now make a rotating-wave approximation to omit the counter-rotating terms and get  
\begin{align} \label{Eq:H_R_RWA}
	&{{H}_R}(t) = {\Omega _{\rm \bf A}}\sum\limits_j {\left( {{e^{i({\mathbf{q}} \cdot {{\mathbf{r}}_{{\rm \bf A},j}} + \phi_R )}}a_{j \uparrow }^\dag {a_{j \downarrow }}{e^{ - i{\omega _R}t}} + {\text{h.c.}}} \right)}  
	+ {\Omega _{\rm 	\bf B}}\sum\limits_j {\left( {{e^{ - i({\mathbf{q}} \cdot {{\mathbf{r}}_{{\rm \bf B},j}} + \phi_R )}}b_{j \uparrow }^\dag {b_{j \downarrow }}{e^{i{\omega _R}t}} + {\text{h.c.}}} \right)} ,
\end{align}
where we have denoted
\begin{subequations}
\begin{align}
&  {\Omega _{\rm \bf A}}{e^{i{\mathbf{q}} \cdot {{\mathbf{r}}_{{\rm \bf A},j}}}} = \frac{{V_R^0}}{4}\int {d{\mathbf{r}}} w_{{\rm \bf A},j \uparrow }^*({\mathbf{r}},t){w_{{\rm \bf A},j \downarrow }}({\mathbf{r}},t){e^{i{\mathbf{q}} \cdot {\mathbf{r}}}}, \\
&  {\Omega _{\rm \bf B}}{e^{ - i{\mathbf{q}} \cdot {{\mathbf{r}}_{{\rm \bf B},j}}}} = \frac{{V_R^0}}{4}\int {d{\mathbf{r}}} w_{ {\rm \bf  B},j \uparrow }^*({\mathbf{r}},t){w_{{\rm \bf B},j \downarrow }}({\mathbf{r}},t){e^{ - i{\mathbf{q}} \cdot {\mathbf{r}}}}. 
\end{align}
\end{subequations}
We then set the amplitudes of the two Raman couplings  ${\Omega _{\rm \bf A}} = {\Omega _{\rm \bf B}} = {\Omega _R}$ and tune the phase factors such that 
\begin{equation}
	{e^{i({\mathbf{q}} \cdot {{\mathbf{r}}_{{\rm \bf A},j}} + \phi_R )}} = 1 \qquad \text{and} \qquad {e^{ - i({\mathbf{q}} \cdot {{\mathbf{r}}_{{\rm \bf B},j}} + \phi_R )}} = -1
\end{equation}
by suitably choosing the phase $\phi_R$ and the momentum $\mathbf{q}$. The Raman term then takes the following form:
\begin{align} \label{Eq:H_R_t}
	&{{ H}_R}(t) = {\Omega _R}\sum\limits_j {\left( {a_{j \uparrow }^\dag {a_{j \downarrow }}{e^{ - i{\omega _R}t}} + {\text{h.c.}}} \right)} 
	 - {\Omega _R}\sum\limits_j {\left( {b_{j \uparrow }^\dag {b_{j \downarrow }}{e^{i{\omega _R}t}} + {\text{h.c.}}} \right)} \;.
\end{align}
By means of the unitary transformation 
\begin{equation} \label{Eq:U_R_t}
	{{ U}_R}(t) = \exp \left\{ { - i\frac{{\omega_R t}}{2}\sum\limits_{j} \left[ {\left( {a_{j \uparrow }^\dag {a_{j \uparrow }} - a_{j \downarrow }^\dag {a_{j \downarrow }}} \right)}  -  {\left( {b_{j \uparrow }^\dag {b_{j \uparrow }} - b_{j \downarrow }^\dag {b_{j \downarrow }}} \right)} \right]} \right\} \,,
\end{equation}
which acts onto the fermionic operators as
\begin{subequations}
	\begin{align}
  &U_R^\dag {a_{j \uparrow }}{U_R} = \exp \left( { - i\frac{{{\omega _R}t}}{2}} \right){a_{j \uparrow }} \; ,\\
  &U_R^\dag {a_{j \downarrow }}{U_R} = \exp \left( {i\frac{{{\omega _R}t}}{2}} \right){a_{j \downarrow }}\; , \\
  &U_R^\dag {b_{j \uparrow }}{U_R} = \exp \left( {i\frac{{{\omega _R}t}}{2}} \right){b_{j \uparrow }} \; , \\
  &U_R^\dag {b_{j \downarrow }}{U_R} = \exp \left( { - i\frac{{{\omega _R}t}}{2}} \right){b_{j \downarrow }} \; ,
\end{align} 
\end{subequations}
the Raman Hamiltonian $H'_R=U^\dagger_R(t) H_R(t) U_R(t)$ becomes time-independent 
\begin{equation} \label{Eq:H_R_final}
	H'_R = \Omega _R\sum\limits_j \left( a_{j \uparrow}^\dag a_{j \downarrow }  - b_{j \uparrow }^\dag b_{j \downarrow } + \text{h.c.} \right).
\end{equation}
We also observe that $H'_R$ does not change if we further apply the unitary transformation ${U}_D(t)$: $H''_R=U^\dagger_D(t) H'_R U_D(t)=H'_R $, i.e., we have
\begin{equation} \label{Eq:H_R_pp}
	H''_R = \Omega _R\sum\limits_j \left( a_{j \uparrow}^\dag a_{j \downarrow }  - b_{j \uparrow }^\dag b_{j \downarrow } + \text{h.c.} \right).
\end{equation}

In the following, we discuss how the Hamiltonians $H'_{\rm IMB}$ and $H'_b$ transform under the transformation $U_R(t)$.
The sublattice imbalance $H'_{\rm IMB}$ becomes 
{\begin{equation} \label{Eq:H_delta}
	H''_{\rm IMB}= \frac{\delta}{2}\sum\limits_{j} \left[ {\left( {a_{j \uparrow }^\dag {a_{j \uparrow }} - a_{j \downarrow }^\dag {a_{j \downarrow }}} \right)}  -  {\left( {b_{j \uparrow }^\dag {b_{j \uparrow }} - b_{j \downarrow }^\dag {b_{j \downarrow }}} \right)} \right] ,
\end{equation}
where $\delta = {\Delta_{\rm IMB}} - \omega_R$ is the two-photon Raman detuning.  Note that even without considering the Raman coupling, the transform $U_R(t)$ in Eq.~\eqref{Eq:U_R_t} can also be made to nullify the large sublattice detuning in $H_{\rm IMB}$ [Eq.~\eqref{Eq:H_IMB}]. For that case, we should replace $\omega_R$ by $\omega_D$. To avoid treating the time-dependent problem with multiple frequency, we might also set them to be commensurate to each other. For simplicity, we just set $\omega_R = \omega_D$. For this case, the two photon Raman detuning $\delta = {\Delta_{\rm IMB}} - \omega_R = {\Delta_{\rm IMB}} - \omega_D$, which can also be understood as the detuning between the sublattice imbalance and the linear shaking frequency, as described in Eq.~(\textcolor{blue}{5}) of the main text.  

When the unitary transformation $U_R(t)$ is applied to the Hamiltonian $H'_b \equiv H'_{\rm NN} + H'_{\rm NNN}$, i.e., $H''_b=U_R^\dagger(t) H'_b U_R(t)$ with $H''_b= H''_{\rm NN} + H''_{\rm NNN}$, we obtain 
{\begin{align}\label{Eq:H_t_final}
 & {H''_{\rm NN}}(t) = \sum\limits_{\left\langle {ij} \right\rangle } {\left( {{t_{ij, \uparrow }}{e^{i\left( {{\mathbf{A}}(t) \cdot {{\mathbf{r}}_{ij}} + {\omega _R}t} \right)}}} \right)a_{i \uparrow }^\dag {b_{j \uparrow }}}  
  +\sum\limits_{\left\langle {ij} \right\rangle } {\left( {{t_{ij, \downarrow }}{e^{i\left( {{\mathbf{A}}(t) \cdot {{\mathbf{r}}_{ij}} - {\omega _R}t} \right)}}} \right)a_{i \downarrow }^\dag {b_{j \downarrow }} + {\text{h.c.}}} , \hfill \\
  &{ H''_{\rm NNN}}(t) = \sum\limits_{\left\langle {\left\langle {ij} \right\rangle } \right\rangle ,s } {\left( {{t_{ij,s }}{e^{i{\mathbf{A}}(t) \cdot {{\mathbf{r}}_{ij}}}}} \right)a_{is }^\dag {a_{js }}}  
   +\sum\limits_{\left\langle {\left\langle {ij} \right\rangle } \right\rangle ,s } {\left( {{t_{ij,s }}{e^{i{\mathbf{A}}(t) \cdot {{\mathbf{r}}_{ij}}}}} \right)b_{is }^\dag {b_{js }} + {\text{h.c.}}} \,. \label{Eq:H_NNN_pp}
\end{align}

Summarizing, the full time-dependent Hamiltonian for the Raman-coupled linearly-shaken spin-dependent hexagonal optical lattice is given by 
\begin{equation}
	 H(t) = H_b + H_{\rm IMB} + H_D(t) + H_R(t), 
\end{equation}
where $H_b$ is the bare hopping term given by Eq.~\eqref{Eq:H_bare}, $H_{\rm IMB}$ is the sublattice imbalance given by Eq.~\eqref{Eq:H_IMB}, $H_D(t)$ is the linear shaking term given by Eq.~\eqref{Eq:H_D_t}, and $H_R(t)$ is the Raman coupling term given by Eq.~\eqref{Eq:H_R_t}. 
After the combined unitary transformation ${U}(t) = {U}_D(t) {U}_R(t)$ [as given by Eqs.~\eqref{Eq:U_D_t} and \eqref{Eq:U_R_t}, respectively], the transformed Hamiltonian can be written as
\begin{equation} \label{Eq:H_t_pp}
	 H''(t) = {H''_b}(t) + {H''_{\rm IMB}} + {H''_R},
\end{equation}
where ${H''_b}(t) = {H''_{\rm NN}}(t) + {H''_{\rm NNN}}(t)$ is the time-dependent hopping term given by Eqs.~\eqref{Eq:H_t_final} and \eqref{Eq:H_NNN_pp}, ${H''_{\rm IMB}}$ is the effective sublattice imbalance given by Eq.~\eqref{Eq:H_delta}, and ${H''_R}$ is the Raman coupling given by Eq.~\eqref{Eq:H_R_pp}. 
We see that the only time-dependent term is the hopping term ${H''_b(t)}$.

\subsubsection{Time-independent effective Hamiltonian}
We now derive a time-independent effective Hamiltonian for the doubly-driven Hamiltonian ${H''_b}(t)$ in Eq.~\eqref{Eq:H_t_pp}, by time-averaging the hopping term within one shaking period $T_D$. To this aim, we make use of the  Jacobi–Anger expansion
\begin{equation}
	{e^{iz\cos \theta }} = \sum\limits_{n =  - \infty }^{ + \infty } i^n {{J_n}(z){e^{in\theta }}},
\end{equation}
where $J_n(z)$ is the $n$-th Bessel function of the first kind.
For the NN coupling, we have
\begin{align} \label{Eq:t_eff_NN}
  &\frac{{t_{ij,s }^{({\text{eff}})}}}{{{t_{ij,s }}}} = {\left\langle {{e^{i\left( {{\mathbf{A}}(t) \cdot {{\mathbf{r}}_{ij}} + {s }{\omega _R}t} \right)}}} \right\rangle _{T_D}} 
    =\frac{{{T_1}}}{T_D}\left[ {{( - 1)}^{{s }}}{e^{ - i{s }\frac{\pi }{2}}}{e^{ - i{z_{ij}}\frac{{T_D - {T_1}}}{T_D}}}{J_{s} }({z_{ij}}) \right. 
-\left.\frac{i}{{2\pi {s }}}{e^{i{z_{ij}}\frac{{{T_1}}}{T_D}}}\left( {{e^{i2\pi {s }\frac{T_D}{{{T_1}}}}} - 1} \right) \right] ,
\end{align}
where ${z_{ij}} = \alpha ({{\mathbf{r}}_{ij}} \cdot {{\mathbf{e}}_F}) $ with $\alpha={F_0}{T_1}/2\pi$.
For the NNN coupling we obtain 
\begin{align} \label{Eq:t_eff_NNN}
	&\frac{{t_{ij,s }^{({\text{eff}})}}}{{{t_{ij,s }}}} ={\left\langle {{e^{i{\mathbf{A}}(t) \cdot {{\mathbf{r}}_{ij}}}}} \right\rangle _{T_D}} =
	 \frac{{{T_1}}}{T_D}\left[ {{e^{ - i{z_{ij}}\frac{{T_D - {T_1}}}{T_D}}}{J_0}({z_{ij}}) + {e^{i{z_{ij}}\frac{{{T_1}}}{T_D}}}\frac{{T_D - {T_1}}}{{{T_1}}}} \right] \,.
\end{align}
In the following, we take $T_1 = T_D/2$  such that $\alpha=F_0T_D/4\pi$, and  express the time-independent effective NN and NNN hopping Hamiltonian as
\begin{align} \label{Eq:H_NN_eff}
&H_{\rm NN}^{({\text{eff}})} = \sum\limits_{\left\langle {ij} \right\rangle } \sum_s { {t_{ij, s }^{({\text{eff}})}} a_{i s }^\dag {b_{j s }}}  +\mathrm{h.c.} \,,  \\
&H_{\rm NNN}^{({\text{eff}})} = \sum\limits_{\left\langle {\left\langle {ij} \right\rangle } \right\rangle} \sum_s {{t_{ij,s }^{({\text{eff}})}} a_{is }^\dag {a_{js }}}  
 +\sum\limits_{\left\langle {\left\langle {ij} \right\rangle } \right\rangle} \sum_s { {t_{ij,s }^{({\text{eff}})}} b_{is }^\dag {b_{js }} + {\text{h.c.}}}
 \label{Eq:H_NNN_eff}
\end{align} 
We then get the time-independent effective Hamiltonian in the real space as follows:
\begin{equation} \label{Eq:H^eff}
	 H^{(\rm eff)} = H_{\rm NN}^{({\text{eff}})}  + H_{\rm NNN}^{({\text{eff}})}  + {H''_{\rm IMB}} + {H''_R},
\end{equation}
where $H_{\rm NN}^{({\text{eff}})}$ ($H_{\rm NNN}^{({\text{eff}})}$) is the effective NN (NNN) hopping term given by Eq.~\eqref{Eq:H_NN_eff} [Eq.~\eqref{Eq:H_NNN_eff}], ${H''_{\rm IMB}}$ is the effective sublattice imbalance given by Eq.~\eqref{Eq:H_delta}, and ${H''_R}$ is the Raman coupling given by Eq.~\eqref{Eq:H_R_pp}.

In order to show that the effective Hamiltonian $H^{(\rm eff)}$ in Eq.~\eqref{Eq:H^eff} can be recast in the form of Eq.~(\textcolor{blue}{1}) of the main text, 
we perform a Fourier transform to the quasimomentum space
\begin{equation}
	{a_{js }} = \sum\limits_{\boldsymbol{k}} {{e^{  i{\boldsymbol{k}} \cdot {{\mathbf{r}}_{{\rm \bf A},j}}}}} {a_{{\boldsymbol{k}}s }},\quad {b_{js }} = \sum\limits_{\boldsymbol{k}} {{e^{  i{\boldsymbol{k}} \cdot {{\mathbf{r}}_{{\rm \bf B},j}}}}} {b_{{\boldsymbol{k}}s }}.
\end{equation}
For the nearest-neighbor hopping term $H_{\rm NN}^{({\text{eff}})} = \sum\limits_{\boldsymbol{k}}H_{\rm NN}^{({\text{eff}})} (\boldsymbol{k})$, we have
\begin{align}
&H_{\rm NN}^{({\text{eff}})} (\boldsymbol{k}) =\sum\limits_{\mathbf{u}} \sum_s {t_{\rm NN,\mathbf{u}}^{({\text{eff}})}(\alpha,\varphi_F){e^{i{\boldsymbol{k}} \cdot {\mathbf{u}}}}} a^\dagger_{\boldsymbol{k}s} b_{\boldsymbol{k}s} + \mathrm{h.c.} \, ,
\end{align}
where
\begin{equation}
	{t_{\rm NN,\mathbf{u}}^{({\text{eff}})}}(\alpha,\varphi_F)= t_{\rm NN}\left(\frac{i}{2}{e^{ - \frac{i}{2}z_{\mathbf{u}}}}{J_1}({z_\mathbf{u}})\right).
\end{equation}
Here, $z_{\mathbf{u}}=\alpha (\mathbf{u} \cdot {{\mathbf{e}}_F}) $, and $\mathbf{u}$ are the three vectors from a sublattice site to its three nearest-neighbor sites. The bare NN hopping constant $t_{\rm NN}$ is assumed to be independent of the hopping direction $\mathbf{u}$. 
We then get
\begin{align} \label{Eq:H_NN_eff_k}
&H_{\rm NN}^{({\text{eff}})} (\boldsymbol{k}) =\sum_s g_{\alpha, \varphi_F}(\boldsymbol{k}) a^\dagger_{\boldsymbol{k}s} b_{\boldsymbol{k}s} + \mathrm{h.c.}\,,
\end{align}
where
\begin{equation}
g_{\alpha, \varphi_F}(\boldsymbol{k}) = \sum\limits_{\mathbf{u}} {t_{\rm NN,\mathbf{u}}^{({\text{eff}})}(\alpha,\varphi_F){e^{i{\boldsymbol{k}} \cdot {\mathbf{u}}}}} 
\end{equation}
is the NN hopping structure factor.

For the NNN term $H_{\rm NNN}^{({\text{eff}})} = \sum\limits_{\boldsymbol{k}}H_{\rm NNN}^{({\text{eff}})} (\boldsymbol{k})$, we obtain 
\begin{align}
&H_{\rm NNN}^{({\text{eff}})} (\boldsymbol{k})=  \sum\limits_{{\mathbf{u'}}} \sum_s {t_{\rm AA,s,{\mathbf{u'}}}^{({\text{eff}})}(\alpha,\varphi_F){e^{i{\boldsymbol{k}} \cdot {\mathbf{u'}}}}} a^\dagger_{\boldsymbol{k}s} a_{\boldsymbol{k}s} 
+\sum\limits_{{\mathbf{u'}}} \sum_s {t_{\rm BB,s,{\mathbf{u'}}}^{({\text{eff}})}(\alpha,\varphi_F){e^{i{\boldsymbol{k}} \cdot {\mathbf{u'}}}}} b^\dagger_{\boldsymbol{k}s} b_{\boldsymbol{k}s} \,.
\end{align} 
Here, the effective hopping constants are given by
\begin{equation}
	{t_{\eta\eta,s,\mathbf{u}' }^{({\text{eff}})}}(\alpha,\varphi_F) = \frac{{{t_{\eta \eta,s }}}}{2}\left[ {{e^{ - \frac{i}{2}{z_{\mathbf{u}'}}}}{J_0}({z_{\mathbf{u}'}}) + {e^{\frac{{i{z_{\mathbf{u}'}}}}{2}}}} \right] ,
\end{equation}
where $z_{{\mathbf{u}'}}=\alpha (\mathbf{u}' \cdot {{\mathbf{e}}_F}) $, and $\mathbf{u}'$ are the six vectors from a sublattice site to its six next-nearest-neighbor sites of the opposite sublattice. Here, we again assume that the bare NNN hopping constants ${t_{\eta \eta,s }}$, which are functions of sublattice and spin indexes, are independent of the hopping directions $\mathbf{u}'$. We note that, for the spin-dependent hexagonal optical lattice discussed above, and as realized experimentally in~\cite{Sengstock20111}, the bare (isotropic) NNN hopping amplitude has the following relationship: 
\begin{equation}
	({t_{AA, \uparrow }} = {t_{BB, \downarrow }}) \gg ({t_{AA, \downarrow }} = {t_{BB, \uparrow }}). 
\end{equation}
For instance, as reported in \cite{Tarnowskii} for the hexagonal lattice with large sublattice imbalance $\Delta_{\rm IMB}$,  ${t_{AA, \uparrow }} \sim (2\pi\times)(80 \text{ to } 115) \,\text{Hz}$, and ${t_{BB, \uparrow }} \sim -(2\pi\times)(2\text{ to }6) \,\text{Hz}$. Thus it is a reasonable approximation to set 
${t_{AA, \downarrow }} = {t_{BB, \uparrow }}=0$ and denote $t_{AA, \uparrow } = t_{BB, \downarrow } \equiv t_{\rm NNN}$. 
We further denote ${t_{\rm AA,\uparrow,\mathbf{u}' }^{({\text{eff}})}}(\alpha,\varphi_F)={t_{\rm BB,\downarrow,\mathbf{u}' }^{({\text{eff}})}}(\alpha,\varphi_F)= {t_{\rm NNN,{\mathbf{u'}}}^{({\text{eff}})}(\alpha,\varphi_F)}$, and then get
\begin{align} \label{Eq:H_NNN_eff_k}
&H_{\rm NNN}^{({\text{eff}})} (\boldsymbol{k})=  g'_{\alpha, \varphi_F}(\boldsymbol{k}) \left(a^\dagger_{\boldsymbol{k}\uparrow} a_{\boldsymbol{k}\uparrow} + b^\dagger_{\boldsymbol{k}\downarrow} b_{\boldsymbol{k}\downarrow} \right) \,,
\end{align} 
where 
\begin{align}
g'_{\alpha, \varphi_F}(\boldsymbol{k}) = \sum\limits_{\mathbf{u}'} {t_{\rm NNN,\mathbf{u}'}^{({\text{eff}})}(\alpha,\varphi_F){e^{i{\boldsymbol{k}} \cdot {\mathbf{u}'}}}} 
\end{align}
is the NNN hopping structure factor. This NNN hopping structure factor is a real function because the summation is performed over six NNN vectors $\mathbf{u}'$ where three of them (with mutual angles $120^\circ$) are exactly the inverse of the other three, and we always have ${t_{\rm NNN,\mathbf{u}'}^{({\text{eff}})}(\alpha,\varphi_F){e^{i{\boldsymbol{k}} \cdot {\mathbf{u}'}}}}  = \left[ {t_{\rm NNN,-\mathbf{u}'}^{({\text{eff}})}(\alpha,\varphi_F){e^{-i{\boldsymbol{k}} \cdot {\mathbf{u}'}}}}  \right]^*$. The hopping terms $H_{\rm NN}^{({\text{eff}})} (\boldsymbol{k})+H_{\rm NNN}^{({\text{eff}})} (\boldsymbol{k}) \equiv H_t(\boldsymbol{k})$ given by Eqs.~\eqref{Eq:H_NN_eff_k} and \eqref{Eq:H_NNN_eff_k} are written as Eq.~(\textcolor{blue}{4}) in the main text. 

In the quasimomentum space,
the spin dependent lattice imbalance term $H''_{\rm IMB} =\sum\limits_{\boldsymbol{k}} H_{\rm IMB} (\boldsymbol{k}) $ gives
{\begin{equation} \label{Eq:H_IMB_k}
	H_{\rm IMB}(\boldsymbol{k})= \frac{\delta}{2}  \left( {a_{ \boldsymbol{k}\uparrow }^\dag {a_{\boldsymbol{k} \uparrow }} - a_{\boldsymbol{k} \downarrow }^\dag {a_{\boldsymbol{k} \downarrow }}} -{b_{\boldsymbol{k} \uparrow }^\dag {b_{\boldsymbol{k} \uparrow }} +b_{\boldsymbol{k} \downarrow }^\dag {b_{\boldsymbol{k}\downarrow }}} \right),
\end{equation}
which is Eq.~(\textcolor{blue}{5}) in the main text;
the Raman term $H''_R =\sum\limits_{\boldsymbol{k}} H_R (\boldsymbol{k}) $  gives rise to
\begin{align} \label{Eq:H_R_k}
	&H_R (\boldsymbol{k})= {\Omega _R} {\left( {a_{\boldsymbol{k} \uparrow }^\dag {a_{\boldsymbol{k}\downarrow }} - {b_{\boldsymbol{k} \uparrow }^\dag {b_{\boldsymbol{k} \downarrow }} + {\text{h.c.}}} }  
 \right)} ,
\end{align} 
which is Eq.~(\textcolor{blue}{7}) in the main text. We then get the time-independent Hamiltonian in the quasimomentum space as $H^{(\text{eff})} = \sum\limits_{\boldsymbol{k}} H^{(\text{eff})}({\boldsymbol{k}})$ with 
\begin{equation}\label{Eq:H_eff_k_final}
 	H^{(\text{eff})}({\boldsymbol{k}}) =  H_{\rm NN}^{({\text{eff}})} (\boldsymbol{k})  + H_{\rm NNN}^{({\text{eff}})} (\boldsymbol{k}) + H_{\rm IMB}(\boldsymbol{k}) + H_R (\boldsymbol{k}),
 \end{equation} 
where the NN hopping, NNN hopping, effective sublattice imbalance, and Raman coupling terms are given by Eqs.~\eqref{Eq:H_NN_eff_k}, \eqref{Eq:H_NNN_eff_k}, \eqref{Eq:H_IMB_k}, and \eqref{Eq:H_R_k}, respectively. 


\subsubsection{Energy spectrum}
To get the energy spectrum for the Hamiltonian $H^{(\text{eff})} = \sum\limits_{\boldsymbol{k}} H^{(\text{eff})}({\boldsymbol{k}})$ as given by Eq.~\eqref{Eq:H_eff_k_final}, we rewrite it as $ H^{(\text{eff})} = \sum\limits_{\boldsymbol{k}} { \psi _{\boldsymbol{k}}^\dag h({\boldsymbol{k}},\tau){\psi _{\boldsymbol{k}}}}$, where ${\psi _{\boldsymbol{k}}} = {\left( {\begin{array}{*{20}{c}}
  {{a_{{\boldsymbol{k}} \uparrow }}}&{{b_{{\boldsymbol{k}} \uparrow }}}&{{a_{{\boldsymbol{k}} \downarrow }}}&{{b_{{\boldsymbol{k}} \downarrow }}} 
\end{array}} \right)^{\rm T}}$ is the $4$-component field operator for the fermions with spin ($s = \uparrow, \downarrow$) and sublattice ($\eta = \rm \bf A, B$) degree-of-freedoms, and $h(\boldsymbol{k},\tau)$ takes the form of Eq.~({\color{blue}{1}}) in the main text. In particular, for the current case, we have
\begin{equation} \label{Eq:h_k_tau_cold-atom}
	h(\boldsymbol{k},\tau) = \left( {\begin{array}{*{20}{c}}
  {\left({d_x({\boldsymbol{k}})},{d_y({\boldsymbol{k}})},{d_z({\boldsymbol{k}},\tau)}\right) \cdot \bm{\sigma} }&{{\Omega _R(\tau)}{\sigma _z}} \\ 
  {{\Omega _R(\tau)}{\sigma _z}}&{\left({d_x({\boldsymbol{k}})},{d_y({\boldsymbol{k}})}, - {d_z({\boldsymbol{k}},\tau)}\right) \cdot \bm{\sigma} } 
\end{array}} \right),
\end{equation}
where the vector $\left({d_x({\boldsymbol{k}})},{d_y({\boldsymbol{k}})},{d_z({\boldsymbol{k}},\tau)}\right)$ is given by
\begin{subequations}
\begin{align}
 {d_x}({\boldsymbol{k}}) &= \operatorname{Re} \left[ {g_{\alpha, \varphi_F}({\boldsymbol{k}})} \right] \,,\\
 {d_y}({\boldsymbol{k}}) &=  - \operatorname{Im} \left[ {g_{\alpha, \varphi_F}({\boldsymbol{k}})} \right] \, ,\\
 {d_z}({\boldsymbol{k}},\tau) &=  [\delta (\tau) + g'_{\alpha, \varphi_F}({\boldsymbol{k}}) ]/2 .
 \end{align}
 \end{subequations}
Here, we have written explicitly the dependence of the ramping parameter $\tau$ for the two-photon Raman detuning $\delta$ and the Raman coupling strength $\Omega_R$. From the expression of Eq.~\eqref{Eq:h_k_tau_cold-atom}, we can read out the Hamiltonian for the system S as ${h_{\rm S}}(\boldsymbol{k},\tau) = \mathbf d_{\rm S}(\boldsymbol k,\tau) \cdot \bm{\sigma}={\left({d_x({\boldsymbol{k}})},{d_y({\boldsymbol{k}})},{d_z({\boldsymbol{k}},\tau)}\right) \cdot \bm{\sigma} }$, the Hamiltonian for the conjugate system as $ {h_{\rm S^*}}(\boldsymbol{k},\tau) = {\left({d_x({\boldsymbol{k}})},{d_y({\boldsymbol{k}})},{-d_z({\boldsymbol{k}},\tau)}\right) \cdot \bm{\sigma} }$, and the coupling between them as $\Lambda(\tau)=\Omega_R(\tau) \sigma_z$. 
In Eq.~\eqref{Eq:h_k_tau_cold-atom}, we have omitted an irrelevant constant energy shift which is proportional to the $4\times4$ identity matrix. 

We can introduce another set of five 4$\times$4 Dirac matrices ${\tilde\Gamma}_\mu$ ($\mu=1,\dots,5$) as follows:
\begin{equation} \label{Eq:Gamma_matrices_tilde}
	{\tilde \Gamma _1} = {s _0} \otimes {\sigma _x},\; {\tilde \Gamma _2} = {s _0} \otimes {\sigma _y},\; {\tilde \Gamma _3} = {s _z} \otimes {\sigma _z}, \;
{\tilde \Gamma _4} = {s _x} \otimes {\sigma _z},\; {\tilde \Gamma _5} = {s _y} \otimes {\sigma _z},
\end{equation}
which also satisfy the following anti-commuting $SO(5)$ Clifford algebra:
\begin{align}
\left\{{\tilde\Gamma}_\mu,{\tilde\Gamma}_\nu\right\}=2\delta_{\mu\nu}. 
\label{eqn:cliff5}
\end{align}
It is direct to see that the Hamiltonian $h(\boldsymbol k,\tau)$ in Eq.~\eqref{Eq:h_k_tau_cold-atom} can be rewritten as
 \begin{align} \label{Eq:h_k_tau_gamma_tilde}
h(\boldsymbol k,\tau)=  d_x(\boldsymbol k) \tilde \Gamma_1 + d_y(\boldsymbol k)\tilde  \Gamma_2 + d_z(\boldsymbol k,\tau) \tilde \Gamma_3 +\Omega_R(\tau) \tilde \Gamma_4,
\end{align}
and the spectrum is again given by
\begin{align}
E_\pm(\boldsymbol k,\tau) = \pm \sqrt{|\boldsymbol d_{\rm S}(\boldsymbol k,\tau)|^2+\Omega_R^2(\tau) }\,.
\end{align}
Finally, we explicitly show the gap $\Delta(\tau)$ with periodic boundary conditions and the gap $\Delta_{\rm edge}(\tau)$, see Eq. \eqref{edgeee}, with armchair edges along the $x$-direction. 
As shown in Fig. \ref{fig_sm2}, the gap $\Delta_{\rm edge}(\tau)$ is finite  at all times during the preparation process and vanishes at the end of the ramp leading to the formation of gapless edge states only in the final target state.

\begin{figure}[htp!]
	\begin{center}
  	\includegraphics[width=0.4\columnwidth]{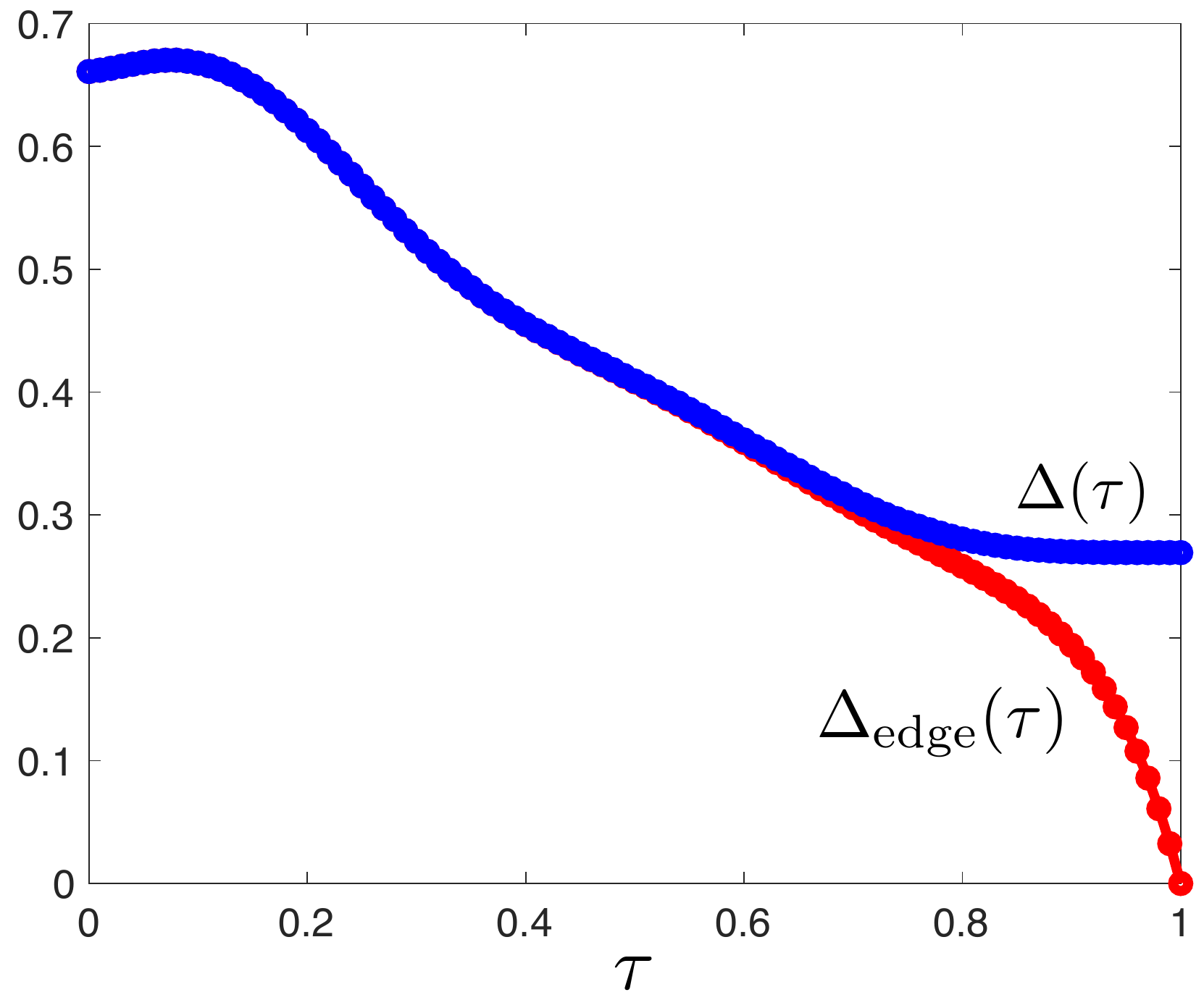}
	\end{center}
	\caption{The gap $\Delta(\tau)$ with periodic boundary conditions and the gap $\Delta_{\rm edge}(\tau)$ with armchair edges as a function of the ramp time $\tau$. 
		The ramping protocol for the detuning term is $\delta(\tau)=\delta_i+(\delta_f-\delta_i)f(\tau)$ with  $f(\tau)=15 \tau^7-70\tau^6+126\tau^5-105 \tau^4+35\tau^3$ and  $\tau=t/T$, with $T$ being the ramp time.
	The coupling between S and S* is $\Lambda(\tau) =\Omega_R(\tau) \sigma_z$ with $\Omega_R(\tau)=\Omega_0 (1-\tau) \sin \pi \tau$. Here: $\delta_i=2$, $\delta_f=0$, $\alpha=3.5$, $\varphi_F=0.09\pi$, $\Omega_0=0.4$, $t_{\rm NNN}=1$ and $t_{\rm NN}=5$. 
	}
	\label{fig_sm2}
\end{figure}

\subsubsection{Vanish of the Chern number for the decoupled total system}
Using $ {\boldsymbol d_{\rm S^*}}(\boldsymbol{k}) = ({d_x},{d_y}, - {d_z})$, it is direct to prove that the Chern numbers for the system and conjugate system are opposite. In particular, taking into account that the Chern number can be expressed as
\begin{align}
C_{\rm S}=\frac{1}{2\pi} \int_{\rm BZ} d {\boldsymbol k} \, \mathcal{F}_{\boldsymbol k, \rm S},
\end{align}
where $\mathcal{F}_{\boldsymbol k, \rm S}$ is the Berry curvature for the system
\begin{align}
\mathcal{F}_{\boldsymbol k, \rm S}=-\frac 12  \mathbf d_{\rm S}(\boldsymbol k) \cdot [ (\partial_{k_x}  \mathbf d_{\rm S}(\boldsymbol k) \times  (\partial_{k_y} \mathbf d_{\rm S}(\boldsymbol k) ]\,,
\end{align}
it is direct to show that the Berry curvature for the conjugate system is
\begin{align}
\mathcal{F}_{\boldsymbol k, \rm S^*}=-\frac 12  \mathbf d_{\rm S^*}(\boldsymbol k) \cdot [ (\partial_{k_x}  \mathbf d_{\rm S^*}(\boldsymbol k) \times  (\partial_{k_y} \mathbf d_{\rm S^*}(\boldsymbol k) ] = -\mathcal{F}_{\boldsymbol k, \rm S} .
\end{align} 
We then get
\begin{equation}
	C_{\rm S^*}=\frac{1}{2\pi} \int_{\rm BZ} d {\boldsymbol k} \, \mathcal{F}_{\boldsymbol k, \rm S^*} = - C_{\rm S},
\end{equation}
and thus the Chern number for the decoupled total system (at $\tau = 0$ and $\tau=1$) vanishes.


\begin{thebibliography}{10}
\bibitem{omegaD}
The detuning $\delta$ can also be understood as the two-photon Raman detuning $\delta = \Delta_\text{IMB} - \omega_R$ [see Fig.~\ref{Fig:Fig2}(b)] as we have set $\omega_D = \omega_R$.
\end{thebibliography}
\end{document}